\pdfoutput=1

\documentclass{jfm}
\usepackage{graphicx}
\usepackage{natbib}
\usepackage{amsmath}

\usepackage{subfigure}

\ifCUPmtlplainloaded \else
  \checkfont{eurm10}
  \iffontfound
    \IfFileExists{upmath.sty}
      {\typeout{^^JFound AMS Euler Roman fonts on the system,
                   using the 'upmath' package.^^J}%
       \usepackage{upmath}}
      {\typeout{^^JFound AMS Euler Roman fonts on the system, but you
                   dont seem to have the}%
       \typeout{'upmath' package installed. JFM.cls can take advantage
                 of these fonts,^^Jif you use 'upmath' package.^^J}%
       \providecommand\upi{\pi}%
      }
  \else
    \providecommand\upi{\pi}%
  \fi
\fi

\ifCUPmtlplainloaded \else
  \checkfont{msam10}
  \iffontfound
    \IfFileExists{amssymb.sty}
      {\typeout{^^JFound AMS Symbol fonts on the system, using the
                'amssymb' package.^^J}%
       \usepackage{amssymb}%
       \let\le=\leqslant  \let\leq=\leqslant
       \let\ge=\geqslant  \let\geq=\geqslant
      }{}
  \fi
\fi

\ifCUPmtlplainloaded \else
  \IfFileExists{amsbsy.sty}
    {\typeout{^^JFound the 'amsbsy' package on the system, using it.^^J}%
     \usepackage{amsbsy}}
    {\providecommand\boldsymbol[1]{\mbox{\boldmath $##1$}}}
\fi

\newcommand\dynpercm{\nobreak\mbox{$\;$dyn\,cm$^{-1}$}}

\usepackage{color}

\title[A Streamwise Constant
Model of Turbulence in Plane Couette Flow]{A Streamwise Constant Model
of Turbulence in Plane Couette Flow}

\author[D. F. Gayme, B. J. McKeon, A. Papachristodoulou, B. Bamieh $\&$ J. C. Doyle ]%
{D.\ns F.\ns G\ls A\ls Y\ls M\ls E$^1$%
\break B.\ns J.\ns M\ls c\ls K\ls
E\ls O\ls N$^1$,\ns A.\ns P\ls A\ls P\ls A\ls C\ls H\ls R\ls
  I\ls S\ls T\ls O\ls D\ls O\ls U\ls L\ls O\ls U$^2$,\break
  B.\ns B\ls A\ls M\ls I\ls E\ls H$^3$ \and J.\ns C.\ns D\ls O\ls Y\ls L\ls E$^1$}

\affiliation{$^1$ Division of Engineering and Applied Science, California Institute of Technology,\\ Pasadena, CA 91106, U.S.A.\\[\affilskip]
$^2$Department of Engineering Science, University of Oxford,
Parks Road, Oxford OX1 3PJ, U.K. \\[\affilskip]
$^3$ Department of Mechanical Engineering, University of California,
Santa Barbara,\\ CA 93106-5070, U.S.A.}

\pubyear{2009} \volume{???} \pagerange{??}
\date{?? and in revised form ??}

\begin{document}

\maketitle

\begin{abstract}
Streamwise and quasi-streamwise elongated structures have been shown
to play a significant role in turbulent shear flows. We model the
mean behavior of fully turbulent plane Couette flow using a
streamwise constant projection of the Navier Stokes equations. This
results in a two-dimensional, three velocity component ($2D/3C$)
model. We first use a steady state version of the model to
demonstrate that its nonlinear coupling provides the mathematical
mechanism that shapes the turbulent velocity profile. Simulations of
the $2D/3C$ model under small amplitude Gaussian forcing of the
cross-stream components are compared to DNS data. The results
indicate that a streamwise constant projection of the Navier Stokes
equations captures salient features of fully turbulent plane Couette
flow at low Reynolds numbers.  A system theoretic approach is used
to demonstrate the presence of large input-output amplification
through the forced $2D/3C$ model. It is this amplification coupled
with the appropriate nonlinearity that enables the $2D/3C$ model to
generate turbulent behaviour under the small amplitude forcing
employed in this study.

\end{abstract}

\section{Introduction}

The Navier Stokes (NS) equations provide a complete dynamical
description of the three velocity components and pressure for simple
canonical flows under the sole modeling assumption that all
important physical phenomena are captured by these equations.
Unfortunately, these infinite dimensional algebraically constrained
equations are analytically intractable. They have however, been
extensively studied computationally and numerical solutions do
exist. For plane Couette flow, the first numerical solution was
computed by \cite{N90}.  \cite{GHC09} provide a detailed discussion
of other work related to a full range of numerical plane Couette
flow solutions.  Ever increasing computing power will continue to
allow progress toward understanding these local properties. However,
a full mathematical understanding of NS even in simple parallel flow
configurations remains elusive, hence considerable effort has been
applied to the search for more analytically-tractable flow models.

In contrast, the Linearized Navier Stokes (LNS) equations can be
analyzed using well developed tools from linear systems theory. For
wall bounded shear flows, one particular property of the LNS that
has been extensively studied is disturbance amplification, e.g.
\cite{F88,G91,RH93,FI93b,BD01,JB05}.  Large disturbance
amplification is common in these flows because the linear operators
governing them are non-normal, (i.e., the operator, $A$, is such
that $A^*A\neq AA^*$).

The LNS are thought to capture the energy production of the full
nonlinear system. \cite{HR94} showed that non-normality and linear
mechanisms are necessary conditions for subcritical transition to
turbulence and it is widely believed that energy amplification is
due to coupling terms that remain in linearized models, see for
example \citep{TTRD93}.   In smooth wall-bounded shear flows the
linear coupling between the Orr-Sommerfeld and Squire equations
associated with nonzero spanwise wave number has been shown to be
required for the generation of the wall layer streaks that are
necessary to maintain turbulence \citep{BF93,KL00}.  In this
context, the term ``streak'' describes the ``well-defined elongated
region of spanwise alternating bands of low and high speed fluid''
\citep{WKH91}. The LNS have also been used by \cite{JB01} to predict
certain second-order statistics of turbulent channel flow. The above
results and a host of others illustrate the power of the LNS as a
model for wall-bounded shear flows.  There is however, one
fundamental flow feature that linear models are unable to capture;
the change in the mean velocity profile as the flow transitions from
laminar to turbulent. In addition, linear analysis can only give
local information regarding the full (nonlinear) system.

Empirical models have been shown to be useful in capturing key
aspects of many flows.  For example, Proper Orthogonal Decomposition
(POD) has been successfully used to construct accurate low
dimensional ordinary differential equation models, e.g.
\citep{L67,SMH05}.  However the preceding analysis utilizes existing
experimental or numerical data, a limitation also applicable to eddy
viscosity models.  In general, data-driven or heuristic models can
be said to suffer from a lack of connection,  of varying degree, to
the governing equations of the problem.

The model studied herein is an attempt to merge the benefits of
studying a physics-based set of equations, such as NS, with the
analytical tractability of a simplified model, such as the LNS.  It
is developed based on the assumption that certain aspects of fully
developed turbulent flow can be reasonably modeled as homogeneous in
the streamwise direction, here denoted ``streamwise constant''. The
idea that a streamwise constant model is sufficient to capture mean
profile changes from laminar to turbulent is strongly supported by
the work of \cite{RI00}, who showed that nonlinear interaction
between the $(k_x, k_z)=(0,\pm N)$ modes, where the $k$'s are the
streamwise and spanwise wave numbers, is the primary factor in
determining the turbulent mean velocity profile in Couette flow.
Further, as was discussed in \cite{OJ94}, this type of model may be
adequate to capture many of the effects associated with the
generation of turbulent wall friction.  A $2\frac{1}{2}$D model
along similar lines has also been developed for the viscous wall
layer; \cite{TP93} for example use such a model to study flow over
riblets in this region.  The physical and analytical basis for
assuming homogeneity in the streamwise direction is discussed in the
following section.

\subsection{Streamwise Coherence}

A growing body of work supports the notion that turbulence in
wall-bounded shear flows is characterized by dynamically significant
coherent structures, particularly features with streamwise and
quasi-streamwise alignment.  Near-wall streaks \citep{KRSR67}, for
example, have been shown to play a key role in energy production
through the `near-wall autonomous cycle' discussed by
\cite{W90,HKW95,W97,JP99}. This cycle is generally agreed to be an
important mechanism in determining the low-order statistics of
turbulent flows in the buffer region and viscous sublayer, i.e.
$y^+\leq 30$ \citep{Schoppa02}.

More recent high Reynolds number studies have focused on the
identification and characterization of streamwise coherence in the
core, i.e. \cite{KA99,MMcJS04,GHA06,HM07_1}.  These motions have
been called large and very large scale motions (respectively LSM and
VLSMs). They appear to have a similar signature to the near-wall
streaks \citep{HM07_2,CMc09}, but tend to be longer in extent, from
one to ten times the outer length scale, $\delta$.  There is
experimental evidence to suggest that at high Reynolds numbers (for
example $Re_{\tau}>7300$), VLSMs contain more energy than the
near-wall structures \citep{MMcJS04,HM07_1,HM07_2}. In turbulent
boundary layers they have also been shown to modulate the near-wall
turbulence, see for example \cite{HM07_2,Mathis09}, suggesting that
they may play an important role in flow dynamics across a range of
scales.

In Couette flow, structures reminiscent of VLSMs have long been
observed in the core through DNS of turbulent plane Couette flow
\citep{LK91,BTAlAn95}.  Although some studies raised the concern
that the structures were numerical artifacts, recent DNS at higher
resolution and with longer box sizes \citep{KLJ96, KawData} have
confirmed the existence of long streamwise alternating high and low
speed streaky structures at the centerline.  In experiments, VLSMs
were first identified through observations of a noticeable peak in
the Fourier energy spectrum of the turbulence intensity at low
frequencies \citep{KLJ96,KU08}.  The Couette flow experiments of
\cite{TA98} found further evidence of very long structures in the
form of long autocorrelations $R_{uu}(\tau)$ and two-point
correlations $R_{uu}(\Delta x)$ as well as periodic variation of
spanwise correlations $R_{uu}(\Delta z)$ in the core. The streamwise
extent of these correlations was longer than those generally seen in
other wall-bounded flows. \cite{KLJ96} also found that in contrast
to other flows, the streamwise correlations for Couette flow are
larger at the center than near the wall.  At channel center the zero
cross distances of $R_{uu}(\tau)$ and $R_{uu}(\Delta x)$ have been
observed to be three times that of the corresponding structures in
Poiseuille flow \citep{KNN05}.  This makes Couette flow an ideal
candidate to test the applicability of a streamwise constant model.

Streamwise constant, $k_x=0$, perturbations to the LNS also produce
the largest input-output response for both laminar
\citep{BD01,JB01,FI93b,FI98} and turbulent \citep{dAJ06} base
velocity profiles.  In addition, streaks of streamwise velocity
naturally arise from the set of initial conditions that produce the
largest energy growth \citep{BF92,FI93a}, namely streamwise
vortices.  Even in linearly unstable flows, studies have shown that
the amplitude of streamwise constant structures can exceed that of
the linearly unstable modes \citep{JB04,G91}.  For channel flows,
\cite{BD01} explicitly showed that streamwise constant disturbances
produce energy growth on the order of $Re^3$ whereas streamwise
varying disturbances grow as a function of $Re^{\frac{3}{2}}$.

In the present work we employ a streamwise constant model based on
the previously discussed experimental and analytical evidence of the
importance of streamwise homogenous features. This so-called
two-dimensional, three (velocity) component, henceforth $2D/3C$,
model for plane Couette flow is simulated under small amplitude
Gaussian forcing.  The results demonstrate the ability of this model
to capture some important features of fully developed turbulent
flow.  In particular, it is demonstrated that: (1) the nonlinear
terms in the $2D/3C$ model capture the momentum redistribution
mechanism involved in creating the shape of the turbulent velocity
profile, (2) a stochastically forced $2D/3C$ model can reproduce the
appropriate turbulent mean velocity profile and Reynolds number
trends, and (3) this model produces amplification of small
disturbances that is consistent with input-output studies of the
LNS.  The work is organized as follows: the next sections of the
paper describe the model and simulation approach. Results and
discussion follow, including a comparison between the model and a
DNS dataset, before final conclusions.

\section{The $2D/3C$ Model}
\label{sec:theModel} The $2D/3C$ model discussed herein is obtained
by setting streamwise ($x$-direction) velocity derivatives in the
full NS equations describing Couette flow to zero
\citep{BobbaThesis}.  This can be thought of as a projection of the
NS into the streamwise constant space. One can explicitly show that
for Couette flow this $2D/3C$ formulation also results in a system
with zero streamwise pressure gradient.

The velocity field is decomposed such that
$\vec{\mathbf{u}}=[U+u^{\prime}_{sw};V+v^{\prime}_{sw},W+w^{\prime}_{sw}]$;
where $U=U(y)=y,\;V=W=0$ is the laminar Couette flow and
$(u^{\prime}_{sw},v^{\prime}_{sw},w^{\prime}_{sw})$ are the
corresponding time-dependent deviations from laminar in the
streamwise constant sense.

\begin{figure}
\centerline{\includegraphics[width=0.5\textwidth,clip]{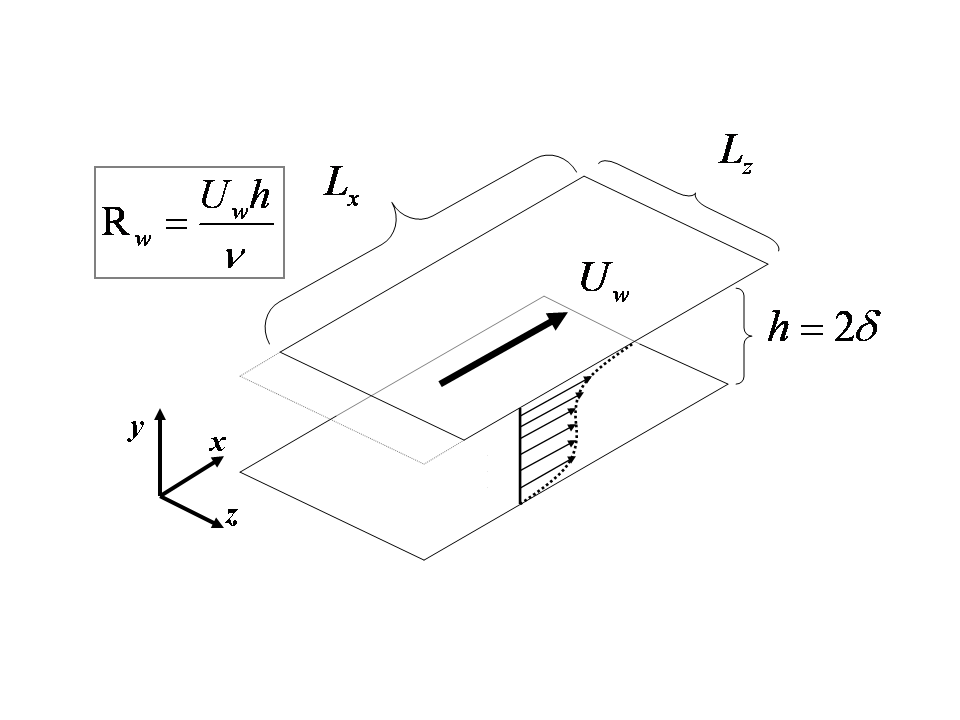}}
\caption{Flow geometry.  Streamwise and spanwise boundaries are
periodic, the bottom wall is stationary and the top wall moves in
the $x$-direction with a velocity $U_w$.  The channel half height is
denoted $\delta$ and the full channel height is denoted $h$.}
\label{fig:flow_geometry}
\end{figure}

The flow geometry is shown in Figure \ref{fig:flow_geometry}. The
Reynolds number employed is $Re_w=\frac{U_w h}{\nu}$, where $U_w$ is
the velocity of the top plate, $h$ is the channel height and $\nu$
is the kinematic viscosity of the fluid.  All distances and
velocities are respectively normalized by $h$ and $U_w$. In the
sequel we will use
$(u^{\prime}_{sw},v^{\prime}_{sw},w^{\prime}_{sw})$ to denote
$\left(\frac{u^{\prime}_{sw}}{U_w},\frac{v^{\prime}_{sw}}{U_w},\frac{w^{\prime}_{sw}}{U_w}\right)$,
and explicitly indicate the scaling only in the figure labels.

A stream function
\[ \label{eqn:streamfunction} v^{\prime}_{sw}=\frac{\partial
\psi}{\partial z}; \;\quad w^{\prime}_{sw}=-\frac{\partial
\psi}{\partial y} \] forces the appropriate $2D$ continuity.  This
results in the following model:
\begin{equation}
\label{eqn:full2D3C}
\begin{aligned}
 \frac{\partial u^{\prime}_{sw} }{\partial t} &=  - \frac{\partial \psi }{\partial z}\frac{\partial u^{\prime}_{sw} }{\partial y}
 - \frac{\partial \psi }{\partial z}\frac{\partial U}{\partial y}
 + \frac{\partial \psi }{\partial y}\frac{\partial u^{\prime}_{sw}}{\partial z} + \frac{1}{Re_w}\Delta u^{\prime}_{sw} \\
 \frac{{\partial \Delta \psi }}{{\partial t}} &=  - \frac{\partial \psi }{\partial z}\frac{\partial \Delta \psi }{\partial y}
 + \frac{\partial \psi }{\partial y}\frac{\partial \Delta \psi }{\partial z} + \frac{1}{Re_w}\Delta ^2 \psi.
  \end{aligned}
\end{equation}
No slip boundary conditions at the wall and periodic boundary
conditions in the spanwise direction are applied (without loss of
generality they can also be used for the stream function equation
since $v^{\prime}_{sw}=w^{\prime}_{sw}=0\Rightarrow \frac{\partial
\psi}{\partial z} =\frac{\partial \psi}{\partial y}= 0
\Rightarrow\psi =\mbox{const}$).  This model retains many of the
important flow features lost in a purely $2D$ model by maintaining
all three velocity components.  Equations (\ref{eqn:full2D3C}) are
an improvement over linear models because it is hypothesized that it
is the nonlinearity in the $u^{\prime}_{sw}(y,z,t)$ equation that
provides the mathematical mechanism for the redistribution of the
fluid momentum. This redistribution creates larger streamwise
velocity gradients in the wall-normal direction and changes the
plane Couette velocity profile from linear to its characteristic
turbulent ``S-shape''. Meanwhile, the important features of the LNS
are maintained.  Linearization of (\ref{eqn:full2D3C}) around the
laminar profile produces a non-normal operator with a coupling term
analogous to the one in the LNS.

The laminar flow solution of Equation (\ref{eqn:full2D3C}) was
previously shown to be globally, that is nonlinearly, stable for all
Reynolds numbers \citep{BBJ02}, and therefore the laminar flow
constitutes a unique solution.  Consequently any transition
mechanisms associated with bifurcations, escape from the basin of
attraction of the laminar solution or the like are not possible. So,
any complications associated with these nonlinear phenomena can be
eliminated from the analysis of these particular equations.  Global
asymptotic stability of the laminar solution also implies that
without forcing, perturbations will eventually decay, in agreement
with the results of \cite{OJ94} who found that after an initial
perturbation a $2D/3C$ model decays (back to laminar) with time. The
fact that one can analytically prove that the unforced $2D/3C$ model
has a unique solution suggests that it is far more analytically
tractable than NS.  We do not pursue analytical studies of the
$2D/3C$ model in the current work, but instead concern ourselves
with showing the applicability of the model in describing important
features of the flow field.  However, the fact that global
statements about these equations can be made implies that future
analytical studies are promising.

As with any model, there are assumptions built into the $2D/3C$
model, and it is important to understand how these relate to the
physical phenomena associated with turbulent flows.  Most obviously,
small scale, three-dimensional turbulent activity, including the
specifics of several structures that are known to exist in the full
flow, is not captured.  While this makes appropriate scaling
relationships more difficult to determine, it does not diminish the
potential of the model for predicting and understanding key aspects
of turbulence in plane Couette flow.  The challenge lies in
extending the $2D/3C$ model to incorporate aspects of the streamwise
variation associated three-dimensional turbulent flow.
\subsection{Modeling Framework}
\label{sec:framework}

No model is a perfect representation of reality.  In addition to
modeling assumptions, parameter errors or external influences on the
system in question are often ignored. Inaccurate parameter estimates
or linearization of a nonlinear system may change the model's
ability to predict behavior.  Environmental conditions that affect
(or disturb) the system may also play an important role in its
dynamics.  This role is not captured by a typical model.

\begin{figure}
\begin{minipage}[b]{0.575\linewidth}
\centering
\includegraphics[width=0.5\textwidth,clip]{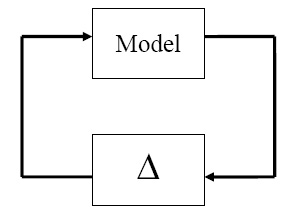}
\caption{A robust control block diagram for a model subject to
uncertainty.  Generally a norm bound on $\Delta$ specifies the
amount of uncertainty that a model can have before a desired
property is lost (i.e., if the model is stable for $\|\Delta\|\leq
1$, this implies robust stability).} \label{fig:robustctrl}
\end{minipage}
\hspace{0.5cm}
\begin{minipage}[b]{0.4\linewidth}
\centering
\includegraphics[width=0.9\textwidth,clip]{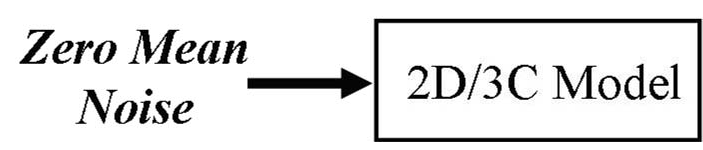}
\caption{The approximation for illustrating the $2D/3C$ model's lack
of robustness.  The zero-mean noise as an approximation for the
modeling errors and uncertainty.  The noise acts as an additive
``uncertainty'' at each time step.} \label{fig:robustctrl_approx}
\end{minipage}
\end{figure}

Robust control theory has historically been used to analyze models
in the presence of such modeling errors (`uncertainty')
\cite{DFT_bk,Z_bk}.  One typically represents all of the
uncertainties using an uncertainty operator $\Delta$.   The block
diagram of Figure \ref{fig:robustctrl} is then used to depict a
model subject to this uncertain set $\Delta$.  Generally robust
control tools provide a bound on $\|\Delta\|$, below which a desired
property can be maintained. Robust control tools do not require a
detailed model of the particular uncertainty. This makes them
appealing in situations where there are unknown (or hard to model)
environmental influences on the system, or when one can only specify
the range on a parameter, rather than an exact value. However, since
the uncertainty is generally specified through a bound that includes
the worst case scenario, the results of this type of analysis may be
very conservative.  One way to mitigate this is to `structure' or
shape the uncertainty, a process which relies on some understanding
of the implicit modeling errors.

In the context of a system comprising a wall-bounded shear flow,
many disturbances can be modeled through the $\Delta$ block in
Figure \ref{fig:robustctrl}.  These sources of modeling errors
(uncertainties) can arise from assumptions on the boundary
conditions or unmodeled dynamics.  External sources of model
uncertainty that are not captured in the NS equations, include
phenomena such as acoustic noise and thermal fluctuations in an
experiment, or the build up of numerical error in simulations. In
addition the uncertain set includes terms excluded by the modeling
assumptions, namely the $k_x\neq 0$ modes in the $2D/3C$ model, or
the nonlinear terms for the LNS model. See \cite{BobbaThesis} for a
full characterization of the types of uncertainties present in shear
flow problems.  Obviously the latter class of perturbations are
strictly bound to satisfy the NS equations, while the former are
less constrained.  Distributed wall roughness (i.e., surface
imperfections present in any real surface), wall vibration,
imperfect alignment of the walls or other parameter estimates may be
captured through either stochastic or other forcing.

In the present work, the framework of robust control is employed in
a nontraditional manner.  Instead of providing an upper bound on
$\|\Delta\|$, (i.e. a robustness guarantee) we describe the extent
to which the laminar flow state is `fragile' (i.e. unable to be
maintained in the face of infinitesimal disturbances). One can think
of this as an inverse robustness (or `fragility') problem, i.e. a
discussion of a lack of robustness. In order to study the
disturbance response of the $2D/3C$ model the system of Figure
\ref{fig:robustctrl} is abstracted into the simplified setting of
Figure \ref{fig:robustctrl_approx}.  We further simplify by
linearizing the $\Delta\psi(y,z,t)$ equation which is equivalent to
recognizing that advection terms in the stream function equation
play a lesser role in redistributing momentum. The forcing and
henceforth $\psi$ are constrained to be small such that the
nonlinear terms are at least an order of magnitude smaller than the
linear ones in all cases studied here. This small amplitude noise
assumption is very important in the development of this work because
of the focus on the effect of small amplitude disturbances on a
fragile system and because larger amplitude forcing can change the
dynamics of the model.

For all of the numerical studies described herein we simulate
\begin{equation}
\label{eqn:linpsi}
\begin{aligned}
 \frac{\partial u^{\prime}_{sw} }{\partial t} &=  - \frac{\partial \psi }{\partial z}\frac{\partial u^{\prime}_{sw} }{\partial y}
 - \frac{\partial \psi }{\partial z}\frac{\partial U}{\partial y}
 + \frac{\partial \psi }{\partial y}\frac{\partial u^{\prime}_{sw}}{\partial z} + \frac{1}{Re_w}\Delta u^{\prime}_{sw}  +d_u\\
 \frac{\partial \Delta \psi}{\partial t}&=\frac{1}{Re_w}\Delta^2 \psi
 +d_\psi,
  \end{aligned}
\end{equation}
with the same boundary conditions as in Equation
(\ref{eqn:full2D3C}).  A short simulation study comparing low-order
streamwise velocity statistics supports the use of linearization in
the $\psi$ equation.

Approximating the full $3D$ system using the interconnection in
Figure \ref{fig:robustctrl} would involve nonlinear mixing modes. In
order to approximate the range of frequencies associated with the
full $3D$ system in the framework of Figure
\ref{fig:robustctrl_approx}, zero mean stochastic forcing was
applied to the $2D/3C$ model.  In particular, the inputs
$d_u(y,z,t)$ and $d_\psi(y,z,t)$ in (\ref{eqn:linpsi}) are small
amplitude and Gaussian, as in \cite{GMcPD09}. The input amplitudes
are defined using the standard deviation, $\sigma_{noise}$. Note
that under these assumptions there is no coupling from the
streamwise components ($u$) back to the cross-stream components (the
$\Delta\psi$ equation).  The plausibility of modeling the type of
disturbances common to experimental conditions in this manner is
confirmed by results from stochastic forcing of the LNS equations,
which leads to flows dominated by streamwise elongated streaks and
vortices that are strikingly similar to those observed in
experiments \citep{FI93b}, as well as by the results of the
simulation study discussed in Section \ref{sec:td}. Further
development of the model would be required to address this effective
feedback mechanism.

\section{Approach}

Time-dependent simulations of the full coupled system
(\ref{eqn:linpsi}) were carried out using a basic second-order
central difference scheme in both the spanwise ($z$) and wall-normal
($y$) directions. Periodic boundary conditions in $z$ and no-slip
boundary conditions in $y$ were applied.  Simulations using the
spectral methods of \cite{WR00} were also performed for comparison.
The pseudospectral simulations employ a Chebyshev interpolant for
the wall-normal direction and a Fourier method for the spanwise
derivatives. The aspect ratio in all of the simulations was greater
than $12$ to $1$ (spanwise to wall-normal) in order to eliminate box
size effects; specifically the
 usual computational box size was $L_y \times L_z = h
\times ~12.8 h$ with $75 \times 100$ grid points.  A spanwise
extent of $12.8h$ was selected to provide a direct comparison to the
full field DNS data from \cite{KawData}.

In this study, the response of the streamwise velocity,
$u^{\prime}_{sw}$, to forcing of the cross-stream velocity
components, $v^{\prime}_{sw}$ and $w^{\prime}_{sw}$, was examined. A
forcing input of zero mean small amplitude Gaussian noise evenly
applied at each $y$--$z$ plane grid point was selected for $d_\psi$.
The other input forcing, $d_u$, was set to zero based on previous
studies of the LNS, which showed that the response to streamwise
body forcing is significantly smaller than the response to
spanwise/wall-normal plane forcing \citep{JB05}. These studies used
an order of magnitude argument to conclude that the difference in
response scales as $\frac{1}{Re^2}$. Furthermore, it is energy
redistribution by streamwise vorticity (i.e. $\Delta \Psi$) that is
thought to be the primary effect governing the shape of the
turbulent velocity profile \citep{HKW95}. The response in the
streamwise velocity component to this forcing may have a nonzero
mean because of the nonlinearity in the $u$ equation.

The different discretization techniques naturally provide a
comparison of different noise forcing distributions.  For example,
the Chebyshev grid results in a higher concentration of noise
forcing near the walls. Throughout the present work it is assumed
that significant numerical errors are not introduced by the methods
of discretization, i.e. the introduction of significant noise arises
only through the $d$ terms of Equation~(\ref{eqn:full2D3C}).

The time evolution of $\Delta\psi$ in Equation (\ref{eqn:linpsi})
can be seen to be a stochastically forced heat equation, i.e. a
linear stochastic partial differential equation which can be solved
analytically (see for example \cite{S07} or \cite{LuoThesis} and the
references therein).  This is not pursued here because a simulation
is a much simpler way to demonstrate the efficacy of the model.  An
exposition on It\^o calculus and Wiener chaos expansions is beyond
the scope of this paper.  Future work may involve pursuing
analytical solutions to both the linear approximation to $\psi$ and
the full nonlinear system (\ref{eqn:full2D3C}).

\section{Results and Discussion}

The results will be divided into three main sections.  We begin by
analyzing the DNS data of \cite{KawData} in the light of the $2D/3C$
model and confirming the extent to which the assumptions of the
$2D/3C$ model can be adduced through this data. Following that, a
time independent version of Equation (\ref{eqn:full2D3C}) is studied
to verify the implicit model filter between $\psi$ and $u$. Finally,
results from full simulations of the time-dependent Equations
(\ref{eqn:linpsi}) are presented and compared to the DNS data.

\subsection{Comparison of DNS data with $2D/3C$ Modeling Assumptions}

Full details of the DNS dataset can be found in \cite{KawData}; a
brief review of key aspects is given here. Three Reynolds numbers
were considered, $Re_w = 3000, 8600$ and $12800$, all with
computational domain size $L_x\times L_y \times L_z=44.8h\times h
\times 12.8h$, $1024\times 96\times 512$ grid points, and a sampling
time ($\frac{tU_w}{L_x}$) of $91$.  The fourth-order finite
difference scheme proposed in \cite{M95} was employed for the $x$
and $z$ directions.  A second-order finite difference method was
used for the $y$ direction.

The friction coefficient, $C_f=9.59\times 10^{-3}$, is somewhat
higher than in other studies, such as \cite{RJ70}. Filling this
friction factor into the relationship developed by~\cite{R59},
\begin{equation}
 \sqrt {\frac{{C_f }}{2}}  = \frac{G}{{\log _{10} \left({\raise0.5ex\hbox{$\scriptstyle 1$}
\kern-0.1em/\kern-0.15em \lower0.25ex\hbox{$\scriptstyle 4$}}Re_w
\right)}}\quad  \mbox{ where }\quad C_f  = \frac{{\tau _w
}}{{\raise0.5ex\hbox{$\scriptstyle 1$} \kern-0.1em/\kern-0.15em
\lower0.25ex\hbox{$\scriptstyle 2$}}\;\rho
\left({\raise0.5ex\hbox{$\scriptstyle 1$} \kern-0.1em/\kern-0.15em
\lower0.25ex\hbox{$\scriptstyle 2$}}U_w\right) ^2 }
\label{eqn:skinfric}
\end{equation}
with $\tau_w$ used to denote shear stress at the wall, leads to an
experimental constant $G=0.199$. Other values reported in the
literature include $G=0.19$ and $G=0.174$ both from \cite{R59} based
on the data of Reichardt and Robertson respectively and $G=0.182$
from the experimental study of~\cite{ElRe82}.

The turbulent mean velocity profiles, turbulence intensities,
Reynolds stresses and budgets of $\overline{u'_iu'_j}$ from this DNS
show good agreement with the experimental results of
\cite{TillmarkThesis} and the spectral DNS study of \cite{KLJ96}
which used a larger box. The two-point correlations in $u$ indicate
that the box lengths used in both the streamwise, $R_{uu}(\Delta
x)$, and spanwise, $R_{uu}(\Delta z)$, directions are sufficient to
eliminate any boundary condition-related spurious effects.

In what follows, a streamwise constant projection of the DNS data is
approximated through a streamwise ($x$) average over the box length,
which will highlight streamwise coherence of the order of the box
length.  The $x$-averaged DNS data is denoted
$\vec{\mathbf{u}_{x_{ave}}}=(u^{\prime}_{x_{ave}}+U(y),v^{\prime}_{x_{ave}},w^{\prime}_{x_{ave}})$
to distinguish it from true streamwise constant data. Time averages
are indicated by an overbar, $\overline{({\cdot})}$.

The ratio of the energy contained in the $x$-averaged DNS to that of
the full field provides a quantitative measure of the extent to
which the DNS data can be approximated as streamwise constant.  For
this comparison the squared $2$-norm is used to approximate the
energy in each $2$-dimensional $x$-averaged velocity component
\begin{equation}
\label{eqn:norm}
\begin{aligned}
\|\beta\|^2=& \int_{\mathcal{Z}} \int_{0}^{1} \beta(y,z)^2\,dy
\;dz\\
\approx& \frac{\Delta z}{2 L_y\;L_z}
\sum_{k=1}^{N_z-1}\left(\sum_{j=1}^{N_y-1}\frac{\Delta
y_{j+1}}{2}\left[\beta^2(y_{j+1},z_{k+1})+\beta^2(y_j,z_{k+1})+\right.\right.\\
&\hspace{6.25cm}+\left.\left.\beta^2(y_{j+1},z_k)+\beta^2(y_j,z_k)\right]\right),
\end{aligned}
\end{equation}
where $\mathcal{Z}$ is the spanwise extent, $\Delta z=z_2-z_1$ is
the spanwise distance between $z$ grid points and trapezoidal
approximations are used for the inhomogeneous $y$ grid.

\begin{table}
\begin{center}
\def~{\hphantom{0}}
\begin{tabular}{ccc}
\hline Component & Total Energy Norm & Percent of Total
Energy\\&$\|\cdot\|$& in $x$-averaged Norm
\\[-3pt]
\hline\hline
$u$ &0.5334& 99.1\\
\hline
$u-U$ &0.1686& 90.2\\
\hline
$v$&0.0279 &19.0\\
\hline
$w$&0.0412& 15.0\\
\hline
\end{tabular}
\caption{Energy content in the $x$-averaged DNS velocity components
at $Re_w=3000$.} \label{table:energies}
\end{center}
\end{table}

Table \ref{table:energies} shows the total energy (based on the full
DNS field at $Re_w=3000$) and the percentage contained in each of
the $x$-averaged velocity components ($u,\,v,\,w$) as well as in the
deviation from laminar (denoted $u-U$).  This latter quantity is
most representative of the energy associated with the differences in
the mean velocity profile for a turbulent versus a laminar flow. The
computations show that $x$-averaged streamwise velocity contains
$99\%$ of the ($u$) energy, whereas the corresponding deviation from
laminar contains $90\%$. As expected, the $x$-averaging results in a
larger loss of information in the spanwise and wall-normal velocity
components.

\begin{figure}
\begin{center}
\includegraphics[width=\textwidth,clip]{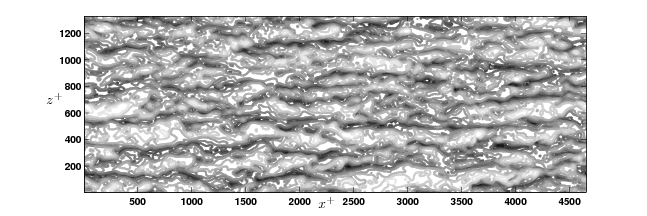}
\caption{An $z$--$x$ plane contour plot of the streamwise velocity,
$u$, from the DNS field, (bottom up view) at $y^+=29$. Light colored
contours denote regions of higher velocity and dark contours
indicate lower velocity regions.} \label{fig:DNS_yplus_cut}
\end{center}
\end{figure}

\begin{figure}
\begin{center}
\subfigure[Average over $25\%$ of streamwise
box]{\label{fig:Contour_u0_Ulam_25percent}
\includegraphics[width=0.45\textwidth,clip]{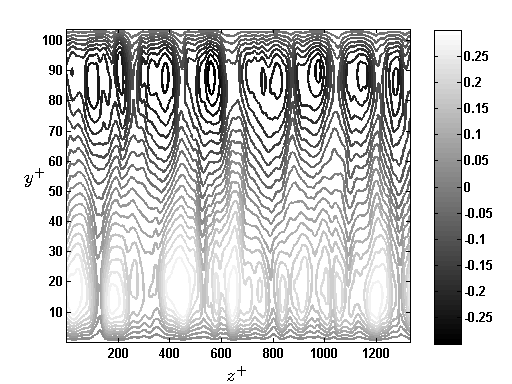}}
\subfigure[Average over full streamwise
box]{\label{fig:Contour_u0_Ulam_full}
\includegraphics[width=0.45\textwidth,clip]{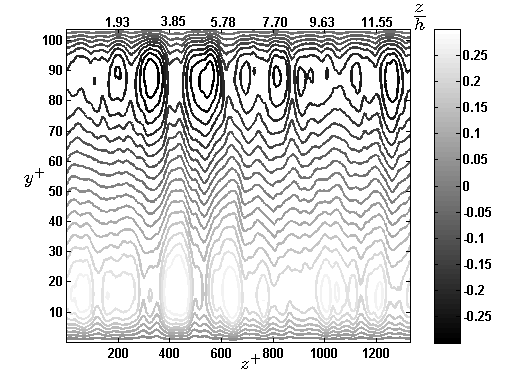}}
\caption{$y$--$z$ plane contour plots of the $x$-averaged (as an
approximation for streamwise constant) DNS deviations from laminar
($u_{{x}_{ave}}^\prime$) (a) averaged over $25\%$ of the streamwise
box length and (b) averaged over the full streamwise box length.}
\label{fig:Contour_u0_Ulam}
\end{center}
\end{figure}

An examination of the DNS streamwise velocity field at $y^+=29$,
close to the outer edge of the region affected by the near-wall
cycle, reveals the signature of streamwise elongated, large scale
streaks in the streamwise/wall-normal plane of the full field
(Figure~\ref{fig:DNS_yplus_cut}).  These streaks are also visible in
Figures \ref{fig:Contour_u0_Ulam_25percent} and
\ref{fig:Contour_u0_Ulam_full} which depict contour plots of the
deviation from laminar flow, $u^{\prime}_{x_{ave}}=u_{x_{ave}}-U$,
when averaged over $25\%$ of the streamwise field and the full field
respectively. Clearly, increasing the averaging length acts as a
filter on structures of different streamwise extent.  The average
over the full box length retains strong evidence of structures
across the entire spanwise/wall-normal plane.  In particular, the
strongest signature near the wall is in qualitative agreement with
the near-wall model of energetic structures centered around
$y^+\approx 15$ with a statistical diameter of $y^+\sim 30$. Another
important feature of Figure \ref{fig:Contour_u0_Ulam_full} is that
the maximum deviations from laminar flow, which are out of spatial
phase with one another, top to bottom, are associated with
large-scale rolling motions which reach across the channel height.

The above analysis shows that there is good agreement between the
DNS data and our assumptions. In the next section this data is used
to suggest a time-independent model for $\psi(y,z)$ in order to
study a steady state version of the streamwise velocity in
(\ref{eqn:full2D3C}).  This is followed by simulation of the full
system (\ref{eqn:linpsi}).

\subsection{Time Independent $u_x(y,z,t)$ Equation}
\label{sec:ss}

The response of Equation \ref{eqn:full2D3C} to a stream function
that is independent of time, $\psi_{ss}(y,z)$, is of interest for
two reasons: (1) the forced solution for the streamwise velocity
permits investigation of whether the $2D/3C$ model filters an
appropriately shaped $\psi_{ss}(y,z)$ towards the expected shape of
the turbulent velocity profile, and (2) it gives insight into the
mathematical mechanisms that create the momentum (energy) transfer
that generates the blunted profile.  The analysis constitutes a
weakly nonlinear analysis, in which the time-independent forcing
takes the form
$\psi_{ss}=\psi_{{ss}_0}+\varepsilon\psi_{{ss}_1}+\dots$.

In \cite{BT07} it was shown that laminar-turbulent flow patterns in
plane Couette flow could be reproduced using a stream function of
the form $\psi(y,z)=\psi_0(y)+\psi_1(y)\cos(k_z z)+\psi_2(y)
\sin(k_z z)$. We use this study as guidance but set the zeroth-order
term $\psi_0$ to zero because a nonzero $\psi_0$ produces a
nonzero-mean spanwise flow $w^\prime_{ss}$, which is not
representative of the velocity field we are interested in studying.
The DNS field \citep{KawData} was also used as a guide to ensure
that the first-order term $\psi_{{ss}_1}$ as well as the
corresponding wall-normal and spanwise velocities, respectively
$v_{{ss}_1}^\prime$ and $w_{{ss}_1}^\prime$, contained
representative features.  For ease of computation and analysis a
simple analytic model for $\psi_{{ss}_1}(y,z)$ was selected, namely
a doubly-harmonic model which matches the boundary conditions
\begin{equation}
\label{eqn:psimodel} \psi_{ss}= \varepsilon\psi_{{ss}_1} (y,z) =
\varepsilon\sin ^2 \left( {\pi y} \right)\cos \left( {\frac{{2\pi
}}{{\lambda_z }}z} \right).
\end{equation}
This corresponds to
\begin{displaymath}
v^\prime_{{ss}_1}(y,z)=-\frac{2\pi }{\lambda_z } \sin ^2 \left( {\pi
y} \right)\sin \left( {\frac{{2\pi}}{{\lambda_z }}z} \right),\mbox{
and } w^\prime_{{ss}_1}(y,z)=-\pi \sin \left( {2\pi y} \right) \cos
\left( {\frac{{2\pi }}{{\lambda_z }}z} \right).
\end{displaymath}
The size of the perturbation, $\varepsilon$, is a free variable to
be explored, while the spanwise wavelength, $\lambda_z$, is fixed to
a value determined using the DNS data.

\begin{figure}
\begin{center}
\subfigure[$\psi_{x_{ave}}$ Estimated from
$w^{\prime}_{x_{ave}}$]{\label{fig:Psi_approx-a}
\includegraphics[width=0.45\textwidth,clip]{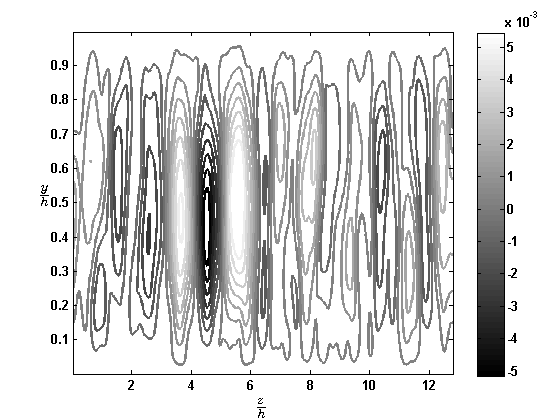}}
\subfigure[$\psi_{ss}(y,z)=0.0675\sin ^2 \left( {\pi y} \right)\cos
\left( {\frac{{2\pi }}{{1.8 }}z} \right)$]{\label{fig:Psi_approx-b}
\includegraphics[width=0.45\textwidth,clip]{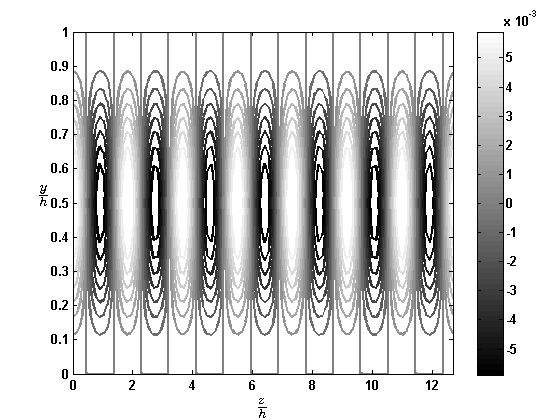}}
\caption{(a) Contour plot of the $x$-averaged DNS (as a streamwise
constant, $2D/3C$ approximation) spanwise velocity deviations
integrated to obtain the stream function,
$\psi_{x_{ave}}(y,z)=-\frac{\partial w^{\prime}_{x_{ave}}}{\partial
y}$. (b) Contour plot of the simple harmonic model for
$\psi_{{ss}}(y,z)=0.00675\sin ^2 \left( {\pi y} \right)\cos \left(
{\frac{{2\pi }}{{1.8 }}z} \right)$ with amplitude and wavelengths
that approximate DNS data.}\label{fig:Psi_approx}
\end{center}
\end{figure}

Streamwise averages of both $v(x,y,z)$ and $w(x,y,z)$ from the DNS
data permit an estimate of $\psi_{{ss}}(y,z)$, (to within some
constant), for that particular field.  A contour plot of the
approximation based on $w^{\prime}_{x_{ave}}(y,z)$ is shown in
Figure $\ref{fig:Psi_approx-a}$.  The value of $\lambda_z\approx 1.8
h$ was chosen to match the results from a Fast Fourier Transform
(FFT) of this data while maintaining the same box size ($12.8h$)
employed for the DNS.  This value is also in the range of the
spanwise wave number corresponding to maximum amplification of the
linear operator (optimal spanwise spacing), $k_z\in~[2.8,4]$
($\lambda_z\in[1.6,2.2]h$), reported in the literature
\citep{FI93b,BF92,G91}.  An initial perturbation amplitude of
$\varepsilon=0.00675$ was selected based on the approximate values
obtained by integrating $v^\prime_{ave}(y,z)$ and
$w^\prime_{ave}(y,z)$.  The estimated amplitude is very small, in
agreement with the idea of using a nominal model plus an uncertainty
$d_\psi$ which is amplified through the coupling in the linear
operator in a manner that is described and quantified in studies
such as \cite{TTRD93} and \cite{JB05}.

A contour plot reflecting these parameter values is provided in
Figure \ref{fig:Psi_approx-b}.  It shows good qualitative agreement,
in particular with the region of strongest signal in the DNS
streamwise average (Figure \ref{fig:Psi_approx-a}). The latter
wall-normal variation is complicated (and Reynolds
number-dependent), but a simple harmonic variation gives a
reasonable representation.  The velocity vector field implied by
Equation (\ref{eqn:psimodel}) is consistent with low speed fluid
being lifted up from the stationary wall and higher speed fluid
being pushed down from the moving wall, and as such supports the
notion that the mechanisms of interest can be modeled using a single
harmonic in both $y$ and $z$.

\begin{figure}
\begin{center}
\subfigure[DNS Data]{\label{fig:SteadyStateContour-a}
\includegraphics[width=0.45\textwidth,clip]{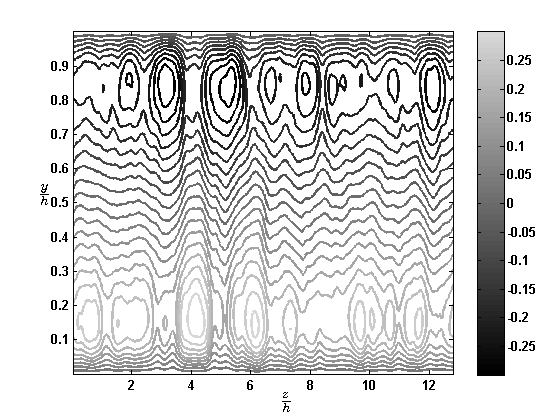}}
\subfigure[$u^{\prime}_{{sw}_{ss}}$ predicted from
$\psi_{ss}$]{\label{fig:SteadyStateContour-b}
\includegraphics[width=0.45\textwidth,clip]{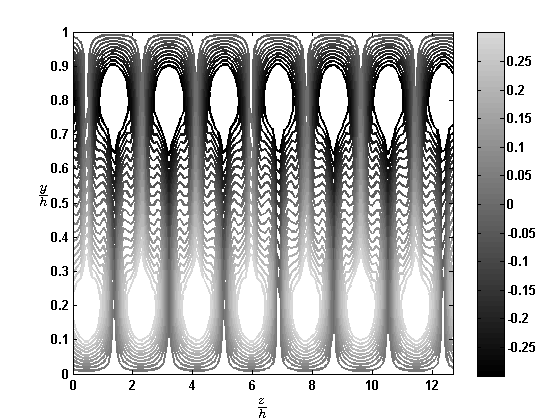}}
\caption{Contour plots of (a) $x$-averaged DNS data and (b) the
$2D/3C$ (streamwise constant) velocity deviations,
$u'_{{ave}_{ss}}$, obtained using steady state estimate
$\psi_{ss}(y,z)=0.00675\sin ^2 \left( {\pi y} \right)\cos \left(
{\frac{{2\pi }}{{1.8143 }}z} \right)$, both plotted with the same
contour levels. }\label{fig:SteadyStateContour}
\end{center}
\end{figure}

\begin{figure}
\begin{center}
\subfigure[$u^{\prime}_{x_{ave}}$ DNS data]
{\label{fig:surfaces-a}\includegraphics[width=0.485\textwidth,clip]{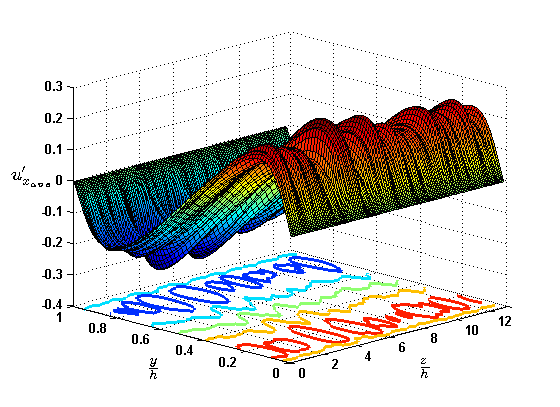}}
\subfigure[$u^{\prime}_{{sw}_{ss}}$ estimated using
$\psi_{ss}(y,z)$]{ \label{fig:surfaces-b}
\includegraphics[width=0.485\textwidth,clip]{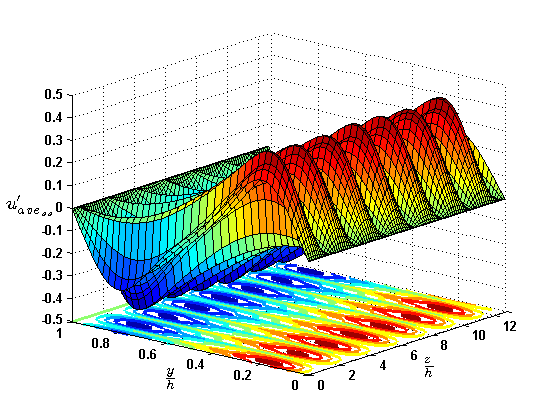}}
\caption{Surface plot of velocity deviations  (a) $u'_{x_{ave}}(y,z)
$ from DNS Data at $Re_w=3000$ and (b)
$u^{\prime}_{{sw}_{ss}}(y,z)$, obtained using steady state estimate
$\psi_{ss}(y,z)=0.00675\sin ^2 \left( {\pi y} \right)\cos \left(
{\frac{{2\pi }}{{1.8143 }}z} \right)$ at $Re_w=3000$, note the $z$
scale difference between (a) and (b)  } \label{fig:surfaces}
\end{center}
\end{figure}

The stream function of Equation \ref{eqn:psimodel} with the selected
$\varepsilon$ and $\lambda_z$ was applied to the time-independent
form of equation \ref{eqn:linpsi}, yielding
\begin{equation}
\label{eqn:ss} \left(- \frac{\partial \psi_{ss} }{\partial
z}\frac{\partial} {\partial y}
 + \frac{\partial \psi_{ss} }{\partial y}\frac{\partial}{\partial z} +
 \frac{1}{Re_w}\Delta\right)
 u^\prime_{{sw}_{ss}}=\frac{\partial \psi_{ss} }{\partial z}\frac{\partial U}{\partial
 y}.
\end{equation}
A contour plot of the resulting $u^{\prime}_{{sw}_{ss}}(y,z)$ is
depicted in Figure \ref{fig:SteadyStateContour}.  This figure shows
a $u^{\prime}_{{sw}_{ss}}(y,z)$ with near-wall rolls that are out of
spanwise phase with one another similar to those seen in the
$x$-averaged DNS data of Figure \ref{fig:Contour_u0_Ulam_full}. The
increased coherence associated with the steady state model relative
to the DNS data manifests as an increased variation in the deviation
from laminar (amplitude of the surface) particularly at the center
of the channel.  This effect is emphasized through comparison of the
surface plots of Figures \ref{fig:surfaces-a} and
\ref{fig:surfaces-b}.  Note the different vertical axis scales for
the two plots.

\begin{figure}
\begin{center}
\subfigure[Amplitude Variation]{\label{fig:SteadyState-a}
\includegraphics[width=0.45\textwidth,clip]{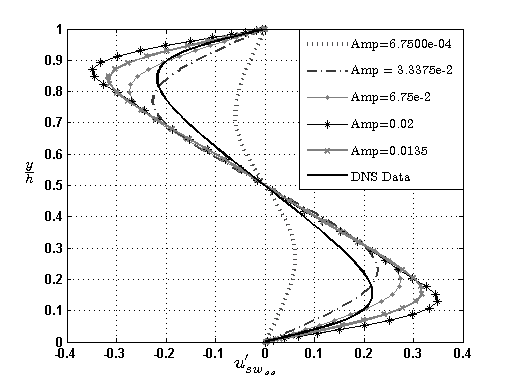}}
\subfigure[Velocity Gradient at the Wall]{\label{fig:SteadyState-b}
\includegraphics[width=0.45\textwidth,clip]{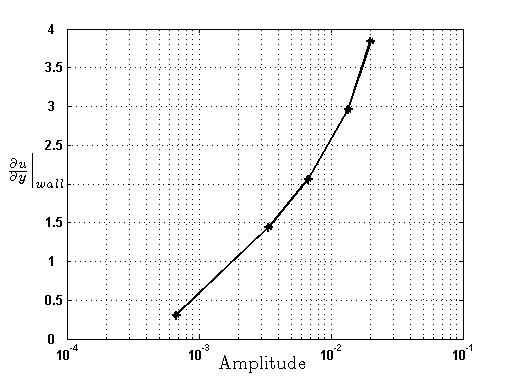}}
\caption{(a) Variation of the $2D/3C$ (streamwise constant) velocity
deviations, $u'_{x_{ave}}$ with $A$; estimates are obtained using
steady state estimate $\psi(y,z)=A \sin ^2 \left( {\pi y}
\right)\cos \left( {\frac{{2\pi }}{{1.8143 }}z} \right)$ (b)
Variation in the velocity gradient at the wall$\left.\frac{\partial
u}{\partial y}\right|_{wall}$ with $\psi_{ss}$
amplitude, $A$.}\label{fig:SteadyState}
\end{center}
\end{figure}

Averages across the span of $u^{\prime}_{sw_{ss}}(y,z)$ for
$\varepsilon=0.00675$ as well as for four additional $\varepsilon$
values are compared to a similar average of $u^{\prime}_{x_{ave}}$
from the DNS in Figure \ref{fig:SteadyState-a}.  Clearly using
$\psi_{ss}$ from (\ref{eqn:psimodel}) as an input to (\ref{eqn:ss})
produces streamwise velocity profiles whose shapes are consistent
with $u^{\prime}_{x_{ave}}\approx \overline{u-U}$ from the DNS. The
peaks are, however located at different wall-normal positions.  An
amplitude that exactly matched both the magnitude and location of
the DNS peaks was not found even when different values of $k_z$ were
studied. This is not unexpected because of the simplicity of the
wall-normal variation in the steady state model, as well as the
streamwise constant and steady state assumptions (clearly the full
turbulent field is neither streamwise constant nor
time-independent). However, this type of model clarifies the
nonlinear role of cross-stream flow features in redistributing
energy in the flow field. These results suggest that the phenomenon
that is responsible for blunting of the velocity profile in the mean
sense is a direct result of the interaction between rolling motions
caused by the $y$--$z$ stream function and the laminar profile.  In
other words, this study provides strong evidence that the
nonlinearity needed to generate the turbulent velocity profile comes
from the nonlinear terms that are present in the
$u^{\prime}_{sw}(y,z,t)$ equation of the $2D/3C$ model
(\ref{eqn:full2D3C}).

The magnitude of forcing applied to the system is reflected in the
amplitude of $\psi_{ss}(y,z)$ which in turn affects the friction
Reynolds number, $Re_{\tau}$ through the skin friction arising due
to the resultant mean velocity gradient at the wall. Increasing the
amplitude ($\varepsilon$) in Equation \ref{eqn:psimodel} is
analogous to increasing the magnitude of the model uncertainty.
These effects are emphasized in Figure \ref{fig:SteadyState-b} which
provides a plot of $\varepsilon$ versus the velocity gradient at the
wall. This behaviour will be discussed in more detail in Section
\ref{sec:td}.

\subsection{Time Dependent $2D/3C$ Model}
\label{sec:td}

Time-dependent simulations of (\ref{eqn:linpsi}) were carried out
using the basic second-order central difference scheme and
pseudo-spectral approaches described earlier.
Table~\ref{table:Reynolds_trends} lists the Reynolds number and
forcing amplitude combinations considered.  The window used for time
averaging was $\Delta t=100000 \frac{h}{U_w}$.

\begin{figure}
\begin{center}
\subfigure[]
{\label{fig:contours2-a}
\includegraphics[width=0.45\textwidth,clip]{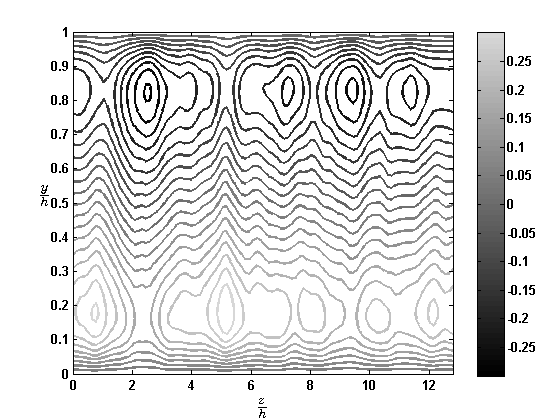}}
\subfigure[]
{\label{fig:contours2-b}
\includegraphics[width=0.45\textwidth,clip]{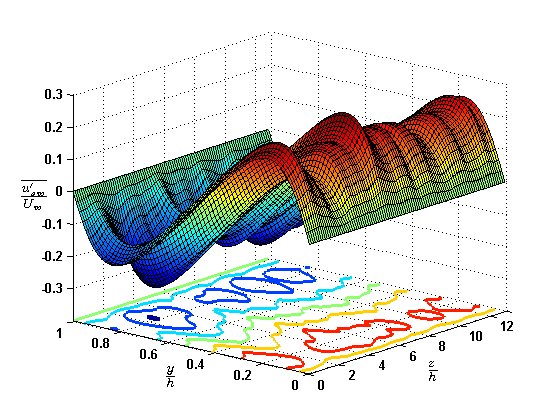}}
\end{center}
\caption{(a) Contour plot of $\overline{u'_{sw}(y,z,t)}$ obtained
from the $2D/3C$ model for Case $1$ with the same contour levels as
in Figure \ref{fig:SteadyStateContour}. (b) The surface plot
corresponding to (a).} \label{fig:contours2}
\end{figure}

The initial simulation (Case $1$, in Table
\ref{table:Reynolds_trends}) was carried out at $Re_w=\frac{U_w
h}{\nu}=3000$ with $d_\psi(x,y,t)$ drawn from a zero mean Gaussian
distribution with standard deviation (noise amplitude)
$\sigma_{noise}=0.01$ applied at every point in the mesh.

\begin{table}
\caption{Computation Details} \centering
\begin{tabular}{cccccc}
\hline Case & Reynolds Number &  $\sigma_{noise}$ &$L_y \times L_z$
&$N_y \times N_z$& Squared Norm of\\&&&&&the Noise Input\\
\hline\hline
1&$3000$ &0.01&$h \times 12.8h$&$75\times 100$&$0.0565$\\
\hline
2&$3000$ &0.0125&$h \times 12.8h$&$75\times 100$&$0.0882$\\
\hline
3&$3000$ &0.004&$h \times 12.8h$&$75\times 100$&$0.009$\\
\hline
4&$8600$&0.004&$h \times 12.8h$&$75\times 130$&$0.0092$\\
\hline
5&$12800$&0.004& $h \times \sim16.5h$&$75\times 130$&$0.0092$\\
\hline
6&$12800$&0.001& $h \times \sim16.5h$&$75\times 130$&$5.77e-04$\\
\hline
Spec 1&$3000$&0.001& $h \times \sim14.5h$&$40\times 81$&$-$\\
\hline
Spec 2&$3000$&0.002& $h \times \sim14.5h$&$40\times 81$&$-$\\
\hline
\end{tabular}
\label{table:Reynolds_trends}
\end{table}

A comparison of Figures \ref{fig:Contour_u0_Ulam_full} and
\ref{fig:contours2-a} shows that contours of constant streamwise
velocity deviation from laminar from the DNS and the $2D/3C$
simulation are in good qualitative agreement. In particular, the
spanwise offset in spatial phase between peaks from top to bottom is
reproduced.  While the dominant wavelength from the $2D/3C$
simulation is somewhat longer than the $\lambda_z\approx 1.8$ of the
DNS (frequency analysis of $\overline{u^{\prime}_{sw}}$ indicates
that most of the energy resides in wave lengths between $4\leq
\lambda_z\leq 6.1$) there is also a significant contribution from
$\lambda_z\approx 2$. There is noticeably better agreement between
the time-dependent model and the DNS (Figure \ref{fig:contours2-b}
and \ref{fig:surfaces-a}) than for the steady state analysis (Figure
\ref{fig:surfaces-b}), likely a consequence of the broadband
stochastic, i.e. less coherent and time-dependent, forcing.

\subsubsection{Mean Velocity Profile}

\begin{figure}
\begin{center}
\subfigure[Mean Velocity Profiles]{\label{fig:Re3000_MeanVelocity-a}
\includegraphics[width=0.475\textwidth,clip]{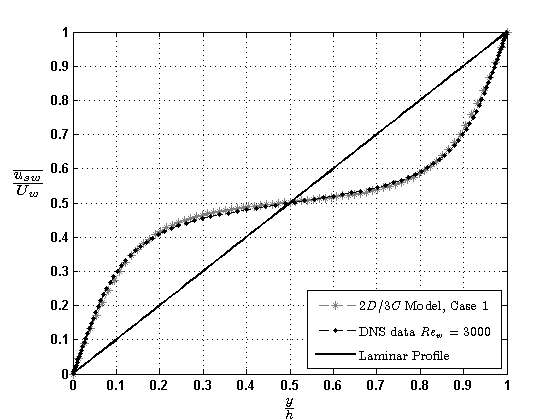}}
\subfigure[Mean Velocity Inner
Units]{\label{fig:Re3000_MeanVelocity-b}
\includegraphics[width=0.475\textwidth,clip]{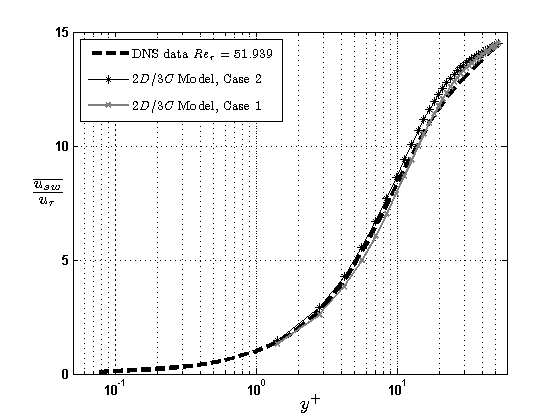}}
\caption{(a) Comparison of mean velocity profile from the $2D/3C$
model Case $1$,  $\overline{u_{sw}(y,z,t)}$, with
$\overline{u(x,y,z,t)}$ from DNS. (b) Inner scaled velocity profiles
comparison of Cases $1$ and $2$ to the DNS data with
$Re_\tau\approx52$ for all data sets.  }
\label{fig:Re3000_MeanVelocity}
\end{center}
\end{figure}

Averaging the streamwise velocity field obtained for Case $1$ in
Table \ref{table:Reynolds_trends} leads to the mean velocity profile
(i.e. $\overline{u_{sw}}$) shown in Figure
\ref{fig:Re3000_MeanVelocity-a}. The mean profile can also be
plotted in inner units (Figure \ref{fig:Re3000_MeanVelocity-b}) with
the use of Equation \ref{eqn:skinfric} (with $G=0.1991$ from
\cite{KawData}) to estimate the friction velocity, $u_\tau$. There
is good agreement between the DNS and the Case $1$ simulation, even
with the assumption of a friction velocity that corresponds to the
full flow.   However, it is clear that below $y^+\approx 20$ the
$2D/3C$ model underestimates the expected velocity profile (maximum
error $7.4\%$), and above that it overshoots it (maximum error
$2.4\%$). There are two obvious first-order interpretations of these
discrepancies.  First, for cases $1$--$6$, the noise is modeled as
being evenly distributed across the grid while in reality the noise
is likely higher in the buffer region due to the proximity of the
wall, and lower in the overlap layer.  An improved noise model might
improve the agreement.  A second interpretation is that further from
the wall the flow is better modeled by the streamwise constant
approximation, while streamwise variation is more important in the
dynamics of the near-wall region (in agreement with the known
variation of the spectral distribution of streamwise energy in the
full flow).

A second (constant) noise amplitude at the same Reynolds number,
Case $2$, is also shown in Figure \ref{fig:Re3000_MeanVelocity-b}.
The agreement with the DNS is certainly improved below $y^+=20$
(maximum error $6.19\%$ at $y^+=19$), but at the expense of larger
error further from the wall ($\sim 5-6\%$ between $20<y^+<30$).
These results further support the idea that a non-uniform noise
forcing with increased noise near the wall versus that at channel
center may more accurately reflect the conditions in a real flow
field.  This idea is further explored in Section
\ref{sec:vary_noise}.

It should be noted that $u_\tau$ can also be computed directly from
the velocity gradient at the wall.  In both cases $Re_\tau$ is
underestimated by around $10\%$ compared to the estimate from
(\ref{eqn:skinfric}).  Because of the limited number of points near
the wall, the friction relationship from the full flow was
preferred, with the understanding that this would only be correct if
the 2D/3C model with $\sigma_{noise}$ exactly reproduced the mean
flow behavior.

\subsubsection{Reynolds Number and Noise Amplitude Trends}

Four additional Reynolds number and $\sigma_{noise}$ amplitudes
pairs were considered.  The details for each of the cases $3$--$6$,
along with the computational domain and spatial resolution, are
provided in Table \ref{table:Reynolds_trends}. Respective values of
the norm $\|\cdot\|^2$, as computed in Equation (\ref{eqn:norm}), of
the noise input computed over the box are also reported, since this
is a more appropriate measure of the forcing when the box size
varies.

\begin{figure}
\begin{center}
\subfigure[]
{\label{fig:Re3000_Re8600_Re12800_trend-a}
\includegraphics[width=0.475\textwidth,clip]{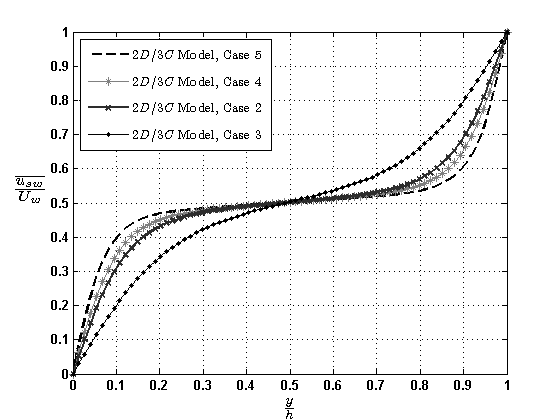}}
\subfigure[]{\label{fig:Re3000_Re8600_Re12800_trend-b}
\includegraphics[width=0.475\textwidth,clip]{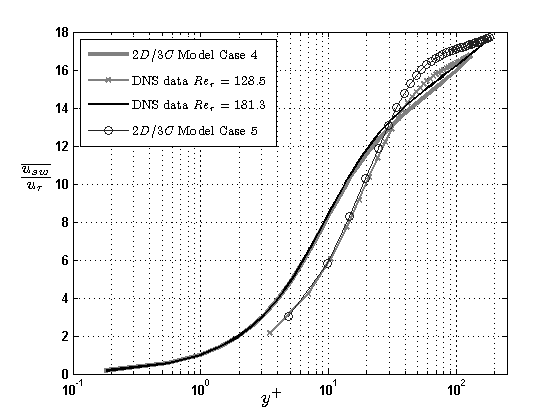}}
\caption{(a) $\overline{u_{sw}(y,z,t)}$ from $2D/3C$ Model for Cases
$2-5$ in Table \ref{table:Reynolds_trends} and (b) a comparison of
$u^+$ versus $y^+$ for Cases $4$ and $5$ with $Re_\tau$ computed
based on the values used in \cite{KawData}.}
\label{fig:Re3000_Re8600_Re12800_trend}
\end{center}
\end{figure}

It is useful to introduce a normalized version of Equation
(\ref{eqn:full2D3C}) through the change of variables $ \tau =
\frac{t}{{Re }} \mbox{ and } \Psi = Re \;\psi$. This creates new
expressions for the forcing in (\ref{eqn:linpsi}),
$\mathcal{D}_u=Re\, d_u (= 0)$ and $\mathcal{D}_\Psi=Re^2 d_\psi$.
The expression $\mathcal{D}_\Psi=Re^2 d_\psi$ indicates that an
increase in noise produces a similar effect to an increase in
Reynolds number (actually $\sqrt{Re}$), as observed in the increased
deviation from laminar observed with increasing noise amplitude in
\ref{fig:SteadyState-a}.  This is especially clear when considering
the variation of $Re_{\tau}$ because an increase in noise amplitude
directly corresponds to increased velocity gradients at the wall due
to the no-slip boundary conditions. Increased profile ``blunting''
with both increasing $\sigma_{noise}$ (noise input energy) and
Reynolds number in cases $2$--$5$ can be observed in figure
\ref{fig:Re3000_Re8600_Re12800_trend-a}.

For the higher Reynolds numbers (but constant noise amplitude) in
Case $4$ and $5$, Figure \ref{fig:Re3000_Re8600_Re12800_trend-b}
shows a worsening agreement in inner units with the DNS data from
\cite{KawData} at $Re_\tau=128.5$ and $Re_\tau=181.3$, respectively.
The underestimation below $y^+\approx 30$ (in the buffer layer) is
more pronounced, but the agreement above $y^+ >30$ remains of
similar magnitude (max error $\sim 4.94\%$ for $Re_\tau=128.5$ and
$8.39\%$ for $Re_\tau=181.3$). We hypothesize that this worsened
agreement may be representative of the increasing scale separation
with increased Reynolds number. Near-wall motions that can
effectively be modeled as streamwise constant at low Reynolds
numbers have an increasingly short streamwise wavelength relative to
the motions that scale with outer length scale $\delta$. That the
zero-error location consistently occurs around $y^+=20$--$30$,
commonly thought to be the upper boundary of the buffer layer, is
consistent with this scale separation argument. For the same reason,
the lack of model resolution in the near-wall region will be
exacerbated with increasing $Re_w$.  In robust control terms, this
points once again to an increase in the model uncertainty near the
wall versus the channel center.  Once again, a better uncertainty
model could be accomplished through the use of a `structured
uncertainty' which would include an increase in $\sigma_{noise}$ in
the near-wall region.

\subsubsection{Varying Noise Amplitude and Distribution}
\label{sec:vary_noise}

\begin{figure}
\begin{center}
\subfigure[]{\label{fig:Re3000_SpectralViscous-a}
\includegraphics[width=0.45\textwidth,clip]{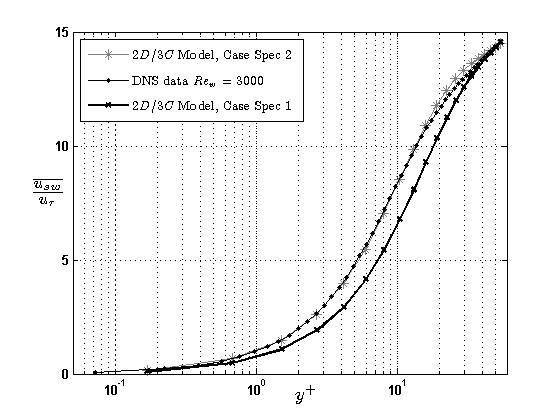}}
\subfigure[]{\label{fig:Re3000_Re12800_same_mean}
\includegraphics[width=0.45\textwidth,clip]{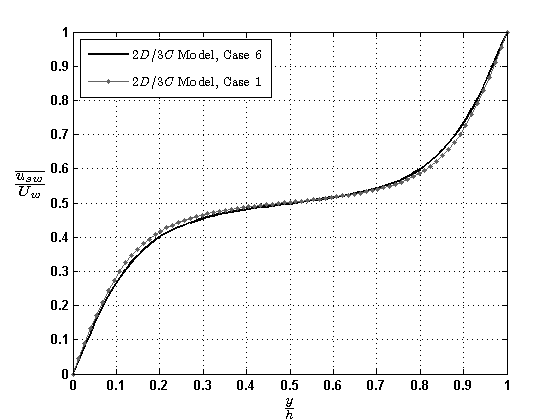}}
\caption{(a) Comparison of $u^+$ versus $y^+$ from $2D/3C$ Model
using Chebyshev spacing in $y$ with DNS data at $Re_w=3000$ based on
$G=0.1991$ ($Re_\tau\approx=52$) (b) Comparison of
$\overline{u_{sw}(y,z,t)}$ from $2D/3C$ Model at for Case $1$
($Re_w=3000$ with $\sigma_{noise}=0.01$, $energy =0.0882$), and Case
$6$ ($Re_w=12800$ with $\sigma_{noise}=0.001$, $energy =5.77e-04$)
same grid and box size.}
\end{center}
\end{figure}

A preliminary effort to introduce a non-uniform distribution of
noise was carried out by repeating the simulation using a
pseudospectral scheme with a Chebyshev interpolant for the
wall-normal direction.  This scheme naturally produces increased
noise near the walls.  Cases Spec $1$ and Spec $2$ in Table
\ref{table:Reynolds_trends} are two such simulations, both at
$Re_w=3000$, with $\sigma_{noise}=0.001$ and $\sigma_{noise}=0.002$
respectively.  Figure \ref{fig:Re3000_SpectralViscous-a} shows the
resulting mean velocity profiles.  Clearly the noise level is too
low for Spec $1$, however for Spec $2$ the maximum error occurs in
the buffer layer and is of the order $5$--$6\%$.  The results of the
spectral simulations indicate that by further noise shaping one
could improve the agreement throughout the profile and across a
range of Reynolds numbers.

As previously discussed there is a strong relationship between the
friction Reynolds number and $\sigma_{noise}$.  As an illustration
of this, Figure \ref{fig:Re3000_Re12800_same_mean} shows that one
can obtain similar mean velocity profiles at two different Reynolds
numbers simply by adjusting the noise amplitude, i.e. a higher
Reynolds number requires a smaller (uniform) noise amplitude to
develop a mean velocity profile that is similar to that of a lower
Reynolds number case with higher noise amplitude.  This result is
consistent with observations of higher transitional Reynolds number
associated with ``quiet'' experiments compared to ones with high
background disturbance levels.  Alternatively, a fixed amplitude
noise produces a larger response (more blunting) at higher $Re$ than
lower ones because disturbance amplification increases with
increasing Reynolds number.  This example makes it clear that the
noise amplitude and the friction Reynolds number are tightly
coupled, while giving further evidence that Reynolds number
dependent wall-normal shaping of the noise would be required to get
a better model representation of the turbulent mean velocity
profiles.

\subsubsection{Characterization of the (small) disturbance amplification}

The results described herein indicate that a very small amount of
stochastic noise forcing limited to the cross-stream components
produces a very large response which corresponds to a behaviour that
is not a solution of the unforced equations.  The ability of our
model (Equation \ref{eqn:full2D3C}), which has a unique solution in
the unforced case, to produce a new flow condition due to such a
forcing supports the notion that the model is not robust to small
disturbances/uncertainty. The potential for disturbance
amplification is not new, in fact it comes directly from the
features of the LNS previously discussed, however the creation and
maintenance of the new flow state is different and cannot come
through the use of a linear model.  A simple characterization of the
amplification maintained through the forced response of
(\ref{eqn:linpsi}), or the lack of robustness (`fragility') of the
system, can be formulated as follows.

Defining the squared $2$-norm of the streamwise component of
(\ref{eqn:linpsi}) (i.e. $\|\overline{u^{\prime}_{sw}}\|^2$) to be
the increase in streamwise energy from the base (laminar) flow, a
so-called amplification factor is given by,
\begin{equation}
\label{eqn:ampfactor}
\Gamma_u=\frac{\|\overline{u^{\prime}_{sw}}\|^2}{\|\sigma_{noise}\|^2}
\end{equation}
which is a measure of the output energy for a given input (noise
forcing amplitude).  $\Gamma_u$ is a nonlinear analog of the
`ensemble energy density' described in previous studies of the
input-output response of the LNS, e.g. \citep{BD01,JB05}.  Those
investigations showed that the coupling between the Orr-Sommerfeld
and Squire modes enables very large, Reynolds number dependent
disturbance amplification.  The amplification factor for case
$3$--$5$, which all have approximately the same input energy, are
respectively $\Gamma_u \approx 680$, $\Gamma_u \approx 2200$, and
$\Gamma_u \approx 2920$. These trends are consistent with the low
Reynolds number scaling trends based on the OSS equations.  This
agreement reflects the effective restriction of the streamwise
constant assumption to amplification of the $k_x=0$ modes (most
amplified in the linear equation) as well as the source of the
amplification in the $2D/3C$ model: a coupling in a similar linear
operator, given by
\begin{equation}
\label{eqn:25Matrix}
\begin{aligned}
\frac{\partial }{\partial t} \left[
\begin{array}{c}
\psi \\
 u_{sw}^{\prime}
\end{array}
\right] = \left[\begin{array}{cc}
\Delta^{-1}\left(\frac{1}{Re_w}\Delta\Delta\right) & 0\\
 -\frac{\partial U}{\partial y} \frac{\partial  }{\partial z} & \frac{1}{Re_w}\Delta
\end{array}\right]
\left[
\begin{array}{c}
\psi \\
 u_{sw}^{\prime}
\end{array}\right].
\end{aligned}
\end{equation}
In this way computing $\Gamma_u$ from the simulation of
(\ref{eqn:linpsi}) is analogous to studying the steady state nonlinear
response to the most amplified $3D$ mode, (i.e. the $k_x=0$ mode).

Equivalent amplification relationships between the cross-stream
velocity components and $\sigma_{noise}$ could similarly be
investigated.

Structures with long streamwise coherence have long been shown to
have a significant role in both transition and fully developed
turbulent flows. Based on these observations we study a streamwise
constant projection of the Navier-Stokes equations for plane Couette
flow, the $2D/3C$ model. Simulation of this model under small
amplitude Gaussian forcing captures the turbulent mean velocity
profile at low Reynolds numbers. Appropriate Reynolds number trends
are also reproduced.

A weakly nonlinear, steady state analysis demonstrates that $2D$
stream functions can produce appropriate mean velocity distributions
when they are nonlinearly coupled to the $2D/3C$ streamwise
velocity.  This indicates that `swirling motions' in the $y$--$z$
plane produce features consistent with the mean characteristics of
fully developed turbulence.  It also provides evidence that the
nonlinear coupling in the $2D/3C$ model is responsible for creating
the well known characteristic `S' shaped turbulent velocity profile.

The use of small amplitude stochastic forcing as an input to the
$2D/3C$ (nominal) model is based on ideas from robust control.
Experimental observations are used to simplify the NS equations to
form this nominal model and the noise forcing is used to capture
both uncertain parameter values and unmodeled effects.  The
resulting forced $2D/3C$ model allows one to isolate phenomena that
can not be decoupled from a full simulation of NS while maintaining
a sufficiently rich description of the physics that govern turbulent
flow.  Our physics-based model should provide greater insight into
the dynamics of the system than an empirical technique.  Such a
model may also allow better control design.

The linearized $2D/3C$ model (\ref{eqn:25Matrix}) maintains the
properties responsible for large disturbance amplification which
have also been linked to subcritical transition. Maintenance of
these linear mechanisms is critical to the success of this approach.
It is the combination of these linear processes along with the
momentum transfer from the two nonlinear terms in the streamwise
velocity equation that enable the model to capture the blunted
turbulent velocity profile. This line of inquiry provides a
complementary perspective to transient growth and structurally based
models, in that the $2D/3C$ model offers some improvement in
analytic tractability at the expense of streamwise detail.  The
results are especially promising because the computational and
analytical tractability of this model makes it well suited to higher
Reynolds number studies.

A natural extension of the present work would be the development of
a more appropriate model for the noise distribution. It is common in
the controls literature for a system to have a so-called structured
uncertainty which is based on the physics of a particular system. In
this work the limitation of noise to only the $\Delta\psi$ equation
represents a first level of such an approach. Knowledge of the
physics, for example that the near-wall region is under-resolved in
the $2D/3C$ model, is a first step. Numerical or experimental
studies aimed at characterizing true spatial noise forcing patterns
would further help in determining the correct model for noise
distribution.

\section{Acknowledgements}

The authors would like to thank H. Kawamura and T. Tsukahara for
providing us with their DNS data.  This research is sponsored in
part through a grant from the Boeing Corporation. B.J.M. gratefully
acknowledges support from NSF-CAREER award number 0747672 (program
managers W. W. Schultz and H. H. Winter).

\bibliographystyle{jfm}
\bibliography{turbulence}

\begin{thebibliography}{58}
\expandafter\ifx\csname natexlab\endcsname\relax\def\natexlab#1{#1}\fi

\bibitem[del \'Alamo \& Jim\'enez(2006)]{dAJ06}
{\sc del \'Alamo, J.~C. \& Jim\'enez, J.} 2006 Linear energy amplification in
  turbulent channels. {\em J. Fluid Mech.\/} {\bf 559}, 205--213.

\bibitem[Bamieh \& Dahleh(2001)]{BD01}
{\sc Bamieh, B. \& Dahleh, M.} 2001 Energy amplification in channel flows with
  stochastic excitation. {\em Phys. of Fluids\/} {\bf 13}~(11), 3258--3269.

\bibitem[Barkley \& Tuckerman(2007)]{BT07}
{\sc Barkley, D. \& Tuckerman, L.~S.} 2007 Mean flow of turbulent-laminar
  patterns in plane {C}ouette flow. {\em J. Fluid Mech.\/} {\bf 576}, 109--137.

\bibitem[Bech {\em et~al.\/}(1995)Bech, Tillmark, Alfredsson \&
  Andersson]{BTAlAn95}
{\sc Bech, K.~H., Tillmark, N., Alfredsson, P.~H. \& Andersson, H.~I.} 1995 An
  investigation of turbulent plane {C}ouette flow at low {R}eynolds numbers.
  {\em J. Fluid Mech.\/} {\bf 286}, 291--325.

\bibitem[Bobba(2004)]{BobbaThesis}
{\sc Bobba, K.~M.} 2004 Robust flow stability: Theory, computations and
  experiments in near wall turbulence. PhD thesis, California Institute of
  Technology, Pasadena, CA, USA.

\bibitem[Bobba {\em et~al.\/}(2002)Bobba, Bamieh \& Doyle]{BBJ02}
{\sc Bobba, K.~M., Bamieh, B. \& Doyle, J.~C.} 2002 Highly optimized
  transitions to turbulence. In {\em Proc. of $41^{st}$ IEEE Conf. on Decision
  and Control\/}, pp. 4559--4562. Las Vegas, NV, USA.

\bibitem[Butler \& Farrell(1992)]{BF92}
{\sc Butler, K.~M. \& Farrell, B.~F.} 1992 Three-dimensional optimal
  perturbations in viscous shear flow. {\em Phys. of Fluids A\/} {\bf 4},
  1637--1650.

\bibitem[Butler \& Farrell(1993)]{BF93}
{\sc Butler, K.~M. \& Farrell, B.~F.} 1993 Optimal perturbations and streak
  spacing in wall-bounded turbulent shear flow. {\em Phys. of Fluids A\/} {\bf
  5}, 774--777.

\bibitem[Chung \& McKeon(2010)]{CMc09}
{\sc Chung, D. \& McKeon, B.~J.} 2010 Large-eddy simulation investigation of
  large-scale structures in a long channel flow. {\em J. Fluid Mech.\/} {\bf
  661}, 341 -- 364.

\bibitem[Doyle {\em et~al.\/}(1991)Doyle, Francis \& Tannenbaum]{DFT_bk}
{\sc Doyle, J.~C., Francis, B. \& Tannenbaum, A.} 1991 {\em Feedback Control
  Theory\/}. New York: MacMillan Company.

\bibitem[El~Telbany \& Reynolds(1982)]{ElRe82}
{\sc El~Telbany, M. M.~M. \& Reynolds, A.~J.} 1982 The structure of turbulent
  plane {C}ouette flow. {\em Trans. ASME: J. of Fluids Engineering\/} {\bf
  104}, 367--372.

\bibitem[Farrell(1988)]{F88}
{\sc Farrell, B.~F.} 1988 Optimal excitation of perturbations in viscous shear
  flow. {\em Phys. of Fluids\/} {\bf 31}~(8), 2093--2102.

\bibitem[Farrell \& Ioannou(1993{\natexlab{{\em a\/}}})]{FI93a}
{\sc Farrell, B.~F. \& Ioannou, P.~J.} 1993{\natexlab{{\em a\/}}} Optimal
  excitation of three-dimensiogal perturbations in viscous constant shear flow.
  {\em Phys. of Fluids A\/} {\bf 5}~(6), 1390--1400.

\bibitem[Farrell \& Ioannou(1993{\natexlab{{\em b\/}}})]{FI93b}
{\sc Farrell, B.~F. \& Ioannou, P.~J.} 1993{\natexlab{{\em b\/}}} Stochastic
  forcing of the linearized {N}avier-{S}tokes equations. {\em Phys. of Fluids
  A\/} {\bf 5}~(11), 2600--2609.

\bibitem[Farrell \& Ioannou(1998)]{FI98}
{\sc Farrell, B.~F. \& Ioannou, P.~J.} 1998 Perturbation structure and spectra
  in turbulent channel flow. {\em Theoretical and Computational Fluid
  Dynamics\/} {\bf 11}, 237--250.

\bibitem[Gayme {\em et~al.\/}(2009)Gayme, McKeon, Papachristodoulou \&
  Doyle]{GMcPD09}
{\sc Gayme, D.~F., McKeon, B., Papachristodoulou, A. \& Doyle, J.~C.} 2009
  $2{D}/3{C}$ model of large scale structures in turbulence in plane {C}ouette
  flow. In {\em Proc. of $6^{th}$ Int'l Symposium on Turbulence and Shear Flow
  Phenomenon\/}. Seoul, Korea.

\bibitem[Gibson {\em et~al.\/}(2009)Gibson, Halcrow \& Cvitanovi\'c]{GHC09}
{\sc Gibson, J.~F., Halcrow, J. \& Cvitanovi\'c, P.} 2009 Equilibrium and
  travelling-wave solutions of plane {C}ouette flow. {\em J. Fluid Mech.\/}
  {\bf 638}, 243--266.

\bibitem[Guala {\em et~al.\/}(2006)Guala, Hommema \& Adrian]{GHA06}
{\sc Guala, M., Hommema, S.~E. \& Adrian, R.~J.} 2006 Large-scale and
  very-large-scale motions in turbulent pipe flow. {\em J. Fluid Mech.\/} {\bf
  554}, 521--542.

\bibitem[Gustavsson(1991)]{G91}
{\sc Gustavsson, L.~H.} 1991 Energy growth of three-dimensional disturbances in
  plane {P}oiseuille flow. {\em J. Fluid Mech.\/} {\bf 224}, 241--260.

\bibitem[Hamilton {\em et~al.\/}(1995)Hamilton, Kim \& Waleffe]{HKW95}
{\sc Hamilton, J.~M., Kim, J. \& Waleffe, F.} 1995 Regeneration mechanisms of
  near-wall turbulence structures. {\em J. Fluid Mech.\/} {\bf 287}, 317--348.

\bibitem[Henningson \& Reddy(1994)]{HR94}
{\sc Henningson, D.~S. \& Reddy, S.~C.} 1994 On the role of linear mechanisms
  in transition to turbulence. {\em Phys. of Fluids\/} {\bf 6}~(3), 1396--1398.

\bibitem[Hutchins \& Marusic(2007{\natexlab{{\em a\/}}})]{HM07_1}
{\sc Hutchins, N. \& Marusic, I.} 2007{\natexlab{{\em a\/}}} Evidence of very
  long meandering structures in the logarithmic region of turbulent boundary
  layers. {\em J. Fluid Mech.\/} {\bf 579}, 1--28.

\bibitem[Hutchins \& Marusic(2007{\natexlab{{\em b\/}}})]{HM07_2}
{\sc Hutchins, N. \& Marusic, I.} 2007{\natexlab{{\em b\/}}} Large-scale
  influences in near-wall turbulence. {\em Phil. Trans. Royal Society London
  A\/} {\bf 365}, 647--664.

\bibitem[Jim\'enez \& Pinelli(1999)]{JP99}
{\sc Jim\'enez, J. \& Pinelli, A.} 1999 The autonomous cycle of near-wall
  turbulence. {\em J. Fluid Mech.\/} {\bf 389}, 335--359.

\bibitem[Jovanovi\'c \& Bamieh(2001)]{JB01}
{\sc Jovanovi\'c, M.~R. \& Bamieh, B.} 2001 The spatio-temporal impulse
  response of the linearized {N}avier-{S}tokes equations. In {\em Proc. of the
  2001 American Control Conf.\/}, pp. 1948--1953. Arlington, VA, USA.

\bibitem[Jovanovi\'c \& Bamieh(2004)]{JB04}
{\sc Jovanovi\'c, M.~R. \& Bamieh, B.} 2004 Unstable modes versus non-normal
  modes in supercritical channel flows. In {\em Proc. of the 2004 American
  Control Conf.\/}, pp. 2245--2250. Boston, MA, USA.

\bibitem[Jovanovi\'c \& Bamieh(2005)]{JB05}
{\sc Jovanovi\'c, M.~R. \& Bamieh, B.} 2005 Componentwise energy amplification
  in channel flows. {\em J. Fluid Mech.\/} {\bf 534}, 145--183.

\bibitem[Kim \& Lim(2000)]{KL00}
{\sc Kim, J. \& Lim, J.} 2000 A linear process in wall-bounded turbulent shear
  flows. {\em Phys. of Fluids Letters\/} {\bf 12}~(8), 1885--1888.

\bibitem[Kim \& Adrian(1999)]{KA99}
{\sc Kim, K.~J. \& Adrian, R.~J} 1999 Very large scale motion in the outer
  layer. {\em Phys. of Fluids\/} {\bf 11}~(2), 417--422.

\bibitem[Kitoh {\em et~al.\/}(2005)Kitoh, Nakabyashi \& Nishimura]{KNN05}
{\sc Kitoh, O., Nakabyashi, K. \& Nishimura, F.} 2005 Study on mean velocity
  and turbulence characteristics of plane {C}ouette flow: Low-{R}eynolds-number
  effects and large longitudinal vortical structure. {\em J. Fluid Mech.\/}
  {\bf 539}, 199--227.

\bibitem[Kitoh \& Umeki(2008)]{KU08}
{\sc Kitoh, O. \& Umeki, M.} 2008 Experimental study on large-scale streak
  structure in the core region of turbulent plane {C}ouette flow. {\em Phys. of
  Fluids\/} {\bf 20}~(2), 025107--025111.

\bibitem[Kline {\em et~al.\/}(1967)Kline, Reynolds, Schraub \&
  Runstadler]{KRSR67}
{\sc Kline, S.~J., Reynolds, W.~C., Schraub, F.~A. \& Runstadler, P.~W.} 1967
  The structure of turbulent boundary layers. {\em J. Fluid Mech.\/} {\bf 30},
  741--773.

\bibitem[Komminaho {\em et~al.\/}(1996)Komminaho, Lundbladh \&
  Johansson]{KLJ96}
{\sc Komminaho, J., Lundbladh, A. \& Johansson, A.~V.} 1996 Very large
  structures in plane turbulent {C}ouette flow. {\em J. Fluid Mech.\/} {\bf
  320}, 259--285.

\bibitem[Lee \& Kim(1991)]{LK91}
{\sc Lee, M.~J. \& Kim, J.} 1991 The structure of turbulence in a simulated
  plane {C}ouette flow. {\em Thin Solid Films\/} {\bf 1}, 531--536.

\bibitem[Lumley(1967)]{L67}
{\sc Lumley, J.~L.} 1967 The structure of inhomogeneous turbulence. In {\em
  Atmospheric Turbulence and Wave Propagation\/} (ed. A.~M.Yaglom \& V.~I.
  Tatarski), pp. 166--178. Moscow: Nauka.

\bibitem[Luo(2006)]{LuoThesis}
{\sc Luo, W.} 2006 Wiener chaos expansion and numerical solutions of stochastic
  partial differential equations. PhD thesis, California Institute of
  Technology, Pasadena, CA, USA.

\bibitem[Mathis {\em et~al.\/}(2009)Mathis, Hutchins \& Marusic]{Mathis09}
{\sc Mathis, R., Hutchins, N. \& Marusic, I.} 2009 Large-scale amplitude
  modulation of the small-scale structures of turbulent boundary layers. {\em
  J. Fluid Mech.\/} {\bf 628}, 311--337.

\bibitem[Morinishi(1995)]{M95}
{\sc Morinishi, Y.} 1995 Conservative properties of finite difference schemes
  for incompressible flow. In {\em Center for Turbulence Research Annual
  Research Briefs\/}, pp. 121–--132. Stanford University/{NASA} Ames.

\bibitem[Morrison {\em et~al.\/}(2004)Morrison, McKeon, Jiang \&
  Smits]{MMcJS04}
{\sc Morrison, J.~F., McKeon, B.~J., Jiang, W. \& Smits, A.~J.} 2004 Scaling of
  the streamwise velocity component in turbulent pipe flow. {\em J. Fluid
  Mech.\/} {\bf 1508}, 99--131.

\bibitem[Nagata(1990)]{N90}
{\sc Nagata, M.} 1990 Three-dimensional finite-amplitude solutions in plane
  {C}ouette flow: bifurcation from infinity. {\em J. Fluid Mech.\/} {\bf 217},
  519--527.

\bibitem[Orlandi \& Jim\'enez(1994)]{OJ94}
{\sc Orlandi, P. \& Jim\'enez, J.} 1994 On the generation of turbulent wall
  friction. {\em Phys. of Fluids\/} {\bf 16}~(2), 634--641.

\bibitem[Reddy \& Ioannou(2000)]{RI00}
{\sc Reddy, S.C. \& Ioannou, P.J.} 2000 {\em Laminar-Turbulent Transition IUTAM
  99\/}, chap. Energy Transfer Analysis of Turbulent Plane {C}ouette Flow, pp.
  211--216. Berlin: Springer-Verlag.

\bibitem[Reddy \& Henningson(1993)]{RH93}
{\sc Reddy, S.~C. \& Henningson, D.~S.} 1993 Energy growth in viscous channel
  flows. {\em J. Fluid Mech.\/} {\bf 252}, 209--238.

\bibitem[Robertson(1959)]{R59}
{\sc Robertson, J.~M.} 1959 On turbulent plane {C}ouette flow. In {\em Proc. of
  the $6^{th}$ Midwestern Conf. on Fluid Mech.\/}, pp. 169–--182.

\bibitem[Robertson \& Johnson(1970)]{RJ70}
{\sc Robertson, J.~M. \& Johnson, H.~F.} 1970 Turbulence structure in plane
  {C}ouette flow. {\em J. Engineering Mech. Division\/} {\bf 96}, 1171--1181.

\bibitem[Schoppa \& Hussain(2002)]{Schoppa02}
{\sc Schoppa, W. \& Hussain, F.} 2002 Coherent structure generation in
  near-wall turbulence. {\em J. Fluid Mech.\/} {\bf 453}, 57--108.

\bibitem[Smith {\em et~al.\/}(2005)Smith, Moehlis \& Holmes]{SMH05}
{\sc Smith, T.~R., Moehlis, J. \& Holmes, P.} 2005 Low-dimensional modelling of
  turbulence using the proper orthogonal decomposition: A tutorial. {\em
  Nonlinear Dynamics\/} {\bf 41}, 275--307.

\bibitem[Swanson(2007)]{S07}
{\sc Swanson, J.} 2007 Variations of the solution to a stochastic heat
  equation. {\em Annals of Probability\/} {\bf 35}, 2122--2159.

\bibitem[Tillmark(1995)]{TillmarkThesis}
{\sc Tillmark, N.} 1995 Experiments on transition and turbulence in plane
  {C}ouette flow. PhD thesis, Royal Institute of Technology, Stockholm, Sweden.

\bibitem[Tillmark \& Alfredsson(1998)]{TA98}
{\sc Tillmark, N. \& Alfredsson, P.~H.} 1998 Large scale structure in turbulent
  plane {C}ouette flow. In {\em Advances in Turbulence {VII}\/} (ed. U.~Frish),
  pp. 59--62. Saint-Jean Cap Ferrat, France: Kluwer, Dordrecht.

\bibitem[Trefethen {\em et~al.\/}(1993)Trefethen, Trefethen, Reddy \&
  Driscoll]{TTRD93}
{\sc Trefethen, L.~N., Trefethen, A.~E., Reddy, S.~C. \& Driscoll, T.~A.} 1993
  Hydrodynamic stability without eigenvalues. {\em Science\/} {\bf 261}~(5121),
  578--584.

\bibitem[Tsukahara {\em et~al.\/}(2006)Tsukahara, Kawamura \& Shingai]{KawData}
{\sc Tsukahara, T., Kawamura, H. \& Shingai, K.} 2006 {DNS} of turbulent
  {C}ouette flow with emphasis on the large-scale structure in the core region.
  {\em J. Turbulence\/} {\bf 7}~(019).

\bibitem[Tullis \& Pollard(1993)]{TP93}
{\sc Tullis, S. \& Pollard, A.} 1993 Modeling the time-dependent flow over
  riblets in the viscous wall region. {\em Applied Scientific Research\/} {\bf
  50}~(3-4), 299--314.

\bibitem[Waleffe(1990)]{W90}
{\sc Waleffe, F.} 1990 Proposal for a self-sustaining mechanism in shear flows.
  {\em Tech. Rep.\/}. Center for Turbulence Research, Stanford
  University/{NASA} Ames.

\bibitem[Waleffe(1997)]{W97}
{\sc Waleffe, F.} 1997 On a self-sustaining process in shear flows. {\em Phys.
  Fluids\/} {\bf 9}~(4), 883--900.

\bibitem[Waleffe {\em et~al.\/}(1991)Waleffe, Kim \& Hamilton]{WKH91}
{\sc Waleffe, F., Kim, J. \& Hamilton, J.} 1991 On the origin of streaks in
  turbulent shear flows. In {\em Eighth Int'l Symposium on Turbulent Shear
  Flows\/}. Munich, Germany.

\bibitem[Weideman \& Reddy(2000)]{WR00}
{\sc Weideman, J.~A.~C. \& Reddy, S.~C.} 2000 A {MATLAB} differentiation matrix
  suite. {\em {ACM} Transactions on Mathematical Software\/} {\bf 26}~(4),
  465--519.

\bibitem[Zhou {\em et~al.\/}(1996)Zhou, Doyle \& Glover]{Z_bk}
{\sc Zhou, K., Doyle, J.~C. \& Glover, K.} 1996 {\em Robust and Optimal
  Control\/}. Upper Saddle River, NJ: Prentice Hall.

\end{thebibliography}

\end{document}